\documentclass[aps,prl,twocolumn,superscriptaddress,floatfix]{revtex4-2}

\usepackage[english]{babel}
\usepackage[dvips]{graphicx}
\usepackage{graphics}
\usepackage{amsfonts}
\usepackage{amsmath}
\usepackage{amssymb}
\usepackage{exscale}
\usepackage{afterpage}
\usepackage{upgreek}
\usepackage{color}	%% FOR COLORED NOTES IN DRAFT. COMMENT THIS LINE FOR FINAL VERSION 
\usepackage{braket}	%% For nice BRAKET notation
\usepackage{multirow}	%% For columns spanning multiple rows in Tables
\usepackage{enumitem}
\usepackage{balance}

\DeclareUnicodeCharacter{2009}{\,} 
\DeclareUnicodeCharacter{2212}{-}

\usepackage{miller}	%% For nice simplified writing of Miller indices

\usepackage[normalem]{ulem}    %% To use \uline for underlining, 
\begin{document}

%%%%%%%%
\title{Electronic structure dimensionality of the quantum-critical ferromagnet YbNi$_4$P$_2$}
%%%%%%%%
\author{J. Dai}
\altaffiliation{These authors contributed equally to this work.}
\affiliation{Universit\'e Paris-Saclay, CNRS,  Institut des Sciences Mol\'eculaires d'Orsay, 
			91405, Orsay, France}
\affiliation{Alba Synchrotron, Carrer de la Llum 2-26
			08290 Cerdanyola del Vall\`{e}s, Barcelona, Spain}
   
\author{A. Antezak}
\altaffiliation{These authors contributed equally to this work.} 
\affiliation{Universit\'e Paris-Saclay, CNRS,  Institut des Sciences Mol\'eculaires d'Orsay, 
			91405, Orsay, France}

\author{W.~Broad}
\affiliation{HH Wills Physics Laboratory, University of Bristol, Tyndall Avenue, Bristol, BS8 1TL, UK}

\author{M.~Thees}
\affiliation{Universit\'e Paris-Saclay, CNRS,  Institut des Sciences Mol\'eculaires d'Orsay, 
			91405, Orsay, France}

\author{V.~Zatko}
\affiliation{Institut Catal\`{a} de Nanoci\`{e}ncia i Nanotecnologia, 
			UAB Campus, Bellaterra (Barcelona) 08193, Spain}

\author{R.~L.~Bouwmeester}
\affiliation{MESA+ Institute for Nanotechnology, University of Twente, 7500 AE Enschede, 
			 The Netherlands} 			
			
\author{F.~Fortuna}
\affiliation{Universit\'e Paris-Saclay, CNRS,  Institut des Sciences Mol\'eculaires d'Orsay, 
			91405, Orsay, France} 

\author{P.~Le~F\`evre}
\affiliation{Synchrotron SOLEIL, L'Orme des Merisiers, Saint-Aubin-BP48, 91192 Gif-sur-Yvette, France}
\affiliation{Univ Rennes, CNRS, IPR - UMR 6251, F-35000 Rennes, France}

\author{J. E. Rault}
\affiliation{Synchrotron SOLEIL, L'Orme des Merisiers, Saint-Aubin-BP48, 91192 Gif-sur-Yvette, France}

\author{K.~Horiba}
\affiliation{Photon Factory, Institute of Materials Structure Science, 
			High Energy Accelerator Research Organization (KEK), 1-1 Oho, Tsukuba 305-0801, Japan}

\author{D.~V.~Vyalikh}
\affiliation{Donostia International Physics Center (DIPC), 20018 Donostia-San Sebastian, Spain}
\affiliation{IKERBASQUE, Basque Foundation for Science, 48013 Bilbao, Spain}

\author{H.~Kumigashira}
\affiliation{Photon Factory, Institute of Materials Structure Science, 
High Energy Accelerator Research Organization (KEK), 1-1 Oho, Tsukuba 305-0801, Japan}

\author{K.~Kliemt}
\affiliation{Physikalisches Institut, Goethe-Universität Frankfurt, Max-von-Laue-Straße 1, 60438 Frankfurt am Main, Germany}

\author{S.~Friedemann}
\affiliation{HH Wills Physics Laboratory, University of Bristol, Tyndall Avenue, Bristol, BS8 1TL, UK}

\author{C.~Krellner}
\affiliation{Physikalisches Institut, Goethe-Universität Frankfurt, Max-von-Laue-Straße 1, 60438 Frankfurt am Main, Germany}

\author{E.~Frantzeskakis}
\email{emmanouil.frantzeskakis@universite-paris-saclay.fr}
\affiliation{Universit\'e Paris-Saclay, CNRS,  Institut des Sciences Mol\'eculaires d'Orsay, 
			91405, Orsay, France}
			
\author{A.~F.~Santander-Syro}
\email{andres.santander-syro@universite-paris-saclay.fr}
\affiliation{Universit\'e Paris-Saclay, CNRS,  Institut des Sciences Mol\'eculaires d'Orsay, 
			91405, Orsay, France} 

\date{\today}
%\pacs{79.60.-i}

% 79.60.-i Photoemission and photoelectron spectra
%%%%%%%%%%%%%%%

%%%%%%%%%%%%%%%
\begin{abstract}
%%%
	YbNi$_4$P$_2$ is the first known ferromagnetic metal showing a second-order
	quantum phase transition. Current theoretical understanding rules out second order ferromagnetic quantum criticality in centrosymmetric 2D and 3D metals. Thus, studying the electronic structure of YbNi$_4$P$_2$ 
	is of prime fundamental importance. 
	Using angle-resolved photoemission spectroscopy, 
	we experimentally prove the existence of 1D Fermi surface contours. 
	In addition, our results demonstrate that part of the electronic structure of 
	YbNi$_4$P$_2$ is made of states of higher dimensionality, 
	thereby bringing into question the fact that ferromagnetic quantum criticality in centrosymmetric crystals,
	is exclusively found in 1D systems. 
	Our experimental data show that the electronic structure of YbNi$_4$P$_2$ 
	is a playground of mixed dimensionality, electron correlations, strong hybridization 
	and spin-orbit coupling, all of them providing new insights in understanding the origin 
	of ferromagnetic quantum criticality.
%%%
\end{abstract}
%%%%%%%%%%%%%%%
%
\maketitle
%%%%%%%%%%%%%%%

%%%%%%%%%%%%%%%%%%%%%%%%%%%%%%%%%%%%%%%%%%%%%%%%%%%%%%%%%%%%%%%%%%%%%%%%%%%%%%%%%%%%%%%
%%%%%%%%%%% INTRODUCTION
%%%%%%%%%%%%%%%%%%%%%%%%%%%%%%%%%%%%%%%%%%%%%%%%%%%%%%%%%%%%%%%%%%%%%%%%%%%%%%%%%%%%%%%

A distinct type of phase transition, referred to as a quantum-critical point (QCP), occurs precisely at absolute zero temperature. At this critical point, thermal excitations cease to manifest, and the phase transition is solely governed by quantum fluctuations stemming from Heisenberg's uncertainty principle \cite{coleman_quantum_2005}. One example of such manifestation is ferromagnetic quantum-criticality (FM QC) and although it has been extensively examined in a theoretical framework, it remains ill understood. The central question is whether or not a continuous ferromagnetic quantum-critical point (FM QCP) exists in clean metals. The current theoretical model to describe FM QC in clean metals with a homogeneous magnetization such as ferromagnets, ferrimagnets and canted ferromagnets, suggests that the coupling of electronic low-energy modes, known as soft modes, to the fluctuations of the order parameter results in a fluctuation-induced first-order transition for systems with a 2D and 3D dimensionality \cite{kirkpatrick_universal_2012}. Thereby, this interaction would make continuous FM QC theoretically impossible in 2D and 3D systems. Thus, only 1D systems are compatible with FM QC in clean metals. 2D or 3D electronic states are in general incompatible with FM QCP and are expected to lead to the above described first order transition, with the sole exception of non-centrosymmetric systems with a very strong spin-orbit interaction \cite{kirkpatrick_ferromagnetic_2020}. Most importantly, the current theoretical framework leaves aside key interaction such as spin-orbit coupling \cite{kirkpatrick_ferromagnetic_2020} and Kondo effect \cite{komijani_model_2018,yamamoto_metallic_2010}, hence, faces limitations in accurately describing the physical properties of quasi-one-dimensional systems hosting the aforementioned properties.

The heavy-fermion material YbNi$_4$P$_2$ is unique 
with respect to FM QC~\cite{steppke_ferromagnetic_2013}. 
It is the stoichiometric metallic ferromagnet with the lowest Curie temperature observed ($T_c = 0.15$ K), hence suggesting its proximity to a QCP. 
In fact, the partial substitution of P by As leads to a systematic reduction of $T_c$, 
and eventually to the destruction of FM order and the occurrence of a 
second order QCP when 10\% of P atoms are substituted~\cite{steppke_ferromagnetic_2013}. 
Moreover, YbNi$_4$P$_2$ is a Kondo lattice with a Kondo temperature $T_K\approx 8$ K 
indicating strong Ni$3d-$Yb$4f$ orbital hybridization, 
according to resistivity and Seebeck coefficient measurements~\cite{krellner_ferromagnetic_2011}.
Thus, YbNi$_4$P$_2$ has risen as a paradigm
of FM QC.

\begin{figure*}
	\centering
        \includegraphics[clip, width=\linewidth]{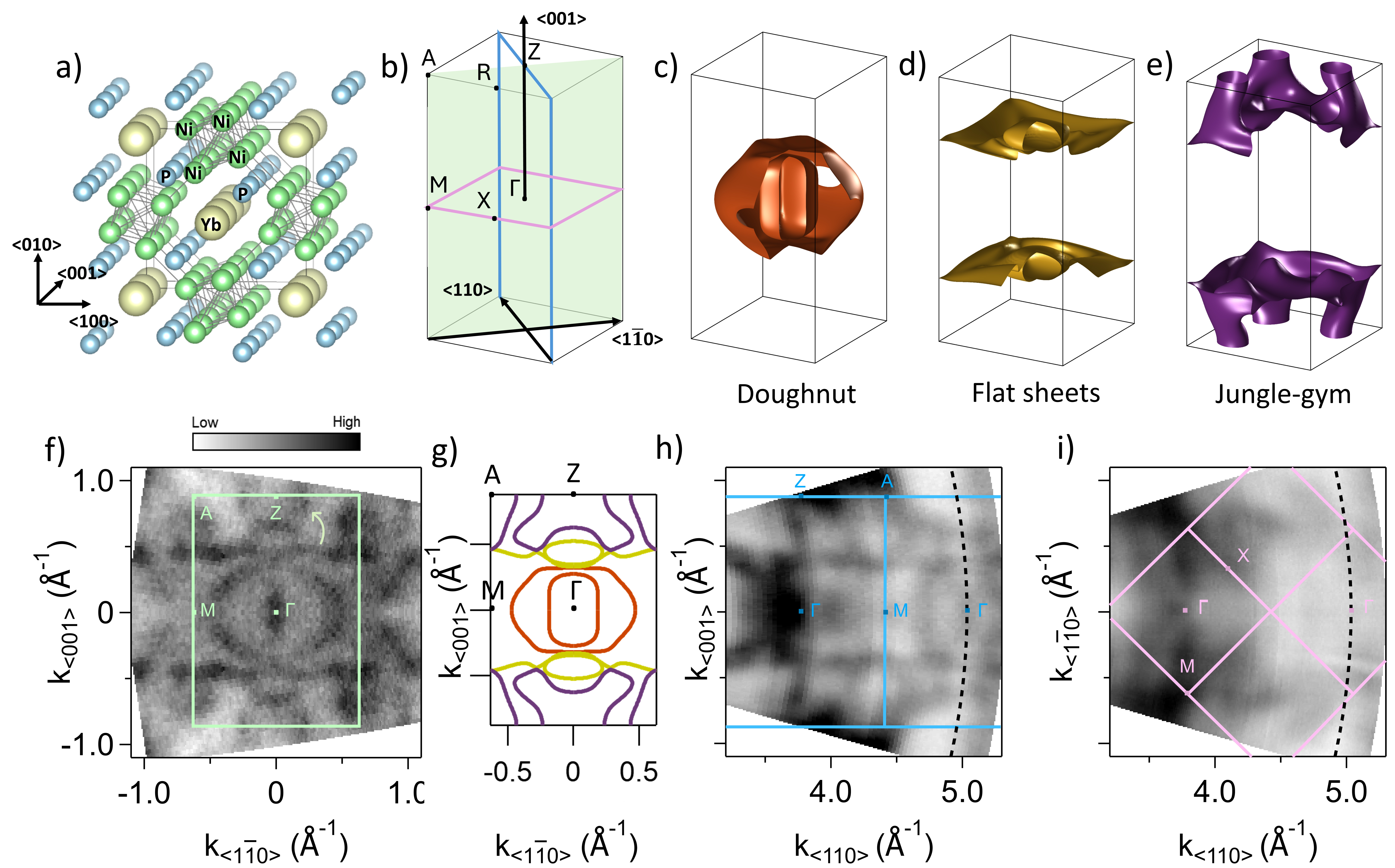}
    \caption{
    		 \footnotesize{
                (a) Crystalline structure of YbNi$_4$P$_2$ with the quasi-1D edge-connected Ni tetrahedra chains along the c axis.
    		(b) Brillouin zone of YbNi$_4$P$_2$ with its high      symmetry points. 
    		The (110)-plane, parallel to the sample surface,      is highlighted in light green. The                      (1$\overline{1}$0) and (001) planes are delimited      by blue and pink contours, respectively.
    		 (c) to (e) DFT constant energy ($E_F - 80$ meV) contours, with the different colors 
    		 corresponding to the bands in Fig. S2 in [SI].
                (f) In-plane constant energy ($E_F - 80$ meV) cut in the 
    		 $\braket{001}\times\braket{1\overline{1}0}$ plane measured with 
    		 linear horizontal light polarization and symmetrized by a mirror operation 
    		 along $k_{\braket{001}} = 0$. 
    		 Symmetrization has been applied to the positive momentum values. 
            (g) Theoretical constant energy ($E_F - 80$ meV) cut. The different colors correspond to the bands in Fig. S2 in [SI].
    		 (h) Out-of-plane constant energy ($E_F - 80$ meV) cut in the 
    		 $\braket{001}\times\braket{110}$ plane measured with 
    		 linear horizontal light polarization.
    		 (i) Out-of-plane constant energy ($E_F - 80$ meV) cut in the  
    		 $\braket{1\overline{1}0}\times\braket{110}$ plane measured with linear horizontal polarized light.
                For both (h) and (i), the horizontal axis corresponds to the direction perpendicular to the surface sample. The scale and tick position of the vertical axes of panels (f)-(i) are all identical. The black dashed line indicates the measurement line in the reciprocal space for a photon energy of 90 eV, used to measure the energy dispersions of Fig. \ref{fig:YNP_Ekmaps}.
    		 }
             %%%
        }
\label{Fig:YNP_FS}
\end{figure*}

In YbNi$_4$P$_2$, Ni is not magnetic~\cite{krellner_ferromagnetic_2011}, 
and the Yb atoms are arranged in chains 
along the $c$-direction and located between edge-connected Ni tetrahedra, 
forming a tetragonal ZrFe$_4$Si$_2$-type structure that is centrosymmetric --see Fig. \ref{Fig:YNP_FS}a. 
Such quasi-one-dimensional crystal structure is reflected in the anisotropy of the resistivity, 
about 5 times larger in-plane than out-of plane at $1.8$ K~\cite{krellner_magnetic_2012}. 
In line with the anisotropic resistivity, band structure calculations have identified Fermi surfaces 
with strong one-dimensional character~\cite{krellner_ferromagnetic_2011}. 
It is believed that such quasi-1D electronic states are at the origin 
of the FM QC in YbNi$_4$P$_2$ \cite{brando_metallic_2016}. 
Hence, although they do not take into account electron correlations or $d-f$ hybridization, 
these theoretical predictions set an appealing basis to understand the physics of this material.
However, an experimental confirmation of quasi-1D Fermi sheets is still missing. 

On one hand, there are so far no direct measurements of the electronic structure 
of YbNi$_4$P$_2$. The main difficulty is that suitable single crystals are difficult to grow and to prepare a clean surface for surface-sensitive spectroscopies. Namely, cleaving is virtually impossible because the crystals are extremely hard, while
no other surface preparation protocols have been established.
On the other hand, while quantum oscillations have been observed \cite{karbassi_anisotropic_2018}, they cannot detect quasi-1D Fermi sheets as the latter do not exhibit any closed Fermi surface contours. Hence, Angle Resolved Photo Emission Spectroscopy (ARPES) is the way forward to understand the complete electronic structure of YbNi$_4$P$_2$ if the surface can be prepared with alternative methods.
Key questions at stake are the very existence of quasi-1D Fermi sheets, 
their interplay or coexistence with Fermi surfaces of higher dimensionality, 
their role in the macroscopic properties of YbNi$_4$P$_2$, 
and the possible consequences of $d-f$ hybridization and electron correlations. 
All these issues are central to the general understanding 
of FM QCP
in heavy-fermion materials~\cite{brando_metallic_2016,komijani_model_2018,chen_continuous_2022}.\\
%%%

%%%

%%%%%%%%%%%%%%%%%%%%%%%%%%
%% HERE WE SHOW
%%%%%%%%%%%%%%%%%%%%%%%%%%
In the present work, we experimentally prove the existence of the quasi-1D contours 
of YbNi$_4$P$_2$ by means of ARPES. 
Thus, our results establish a link between the presence of 1D contours 
and FM QC in this material. 
We observe, however, that these 1D sheets coexist with other constant energy surfaces of 2D and 3D character.
Moreover, the observed electronic structure near the Fermi level ($E_F$) 
shows fingerprints of strong spin-orbit coupling (SOC) and $d - f$ hybridisation, 
demonstrating that these features are important in understanding the properties of YbNi$_4$P$_2$.

%%%%%
% FIGURE 2
%%%%%
%%%%%%%%%%%%%%%%%%%%%%%%%%%%%%%%

\begin{figure}
	\centering
        \includegraphics[clip, width=\linewidth]{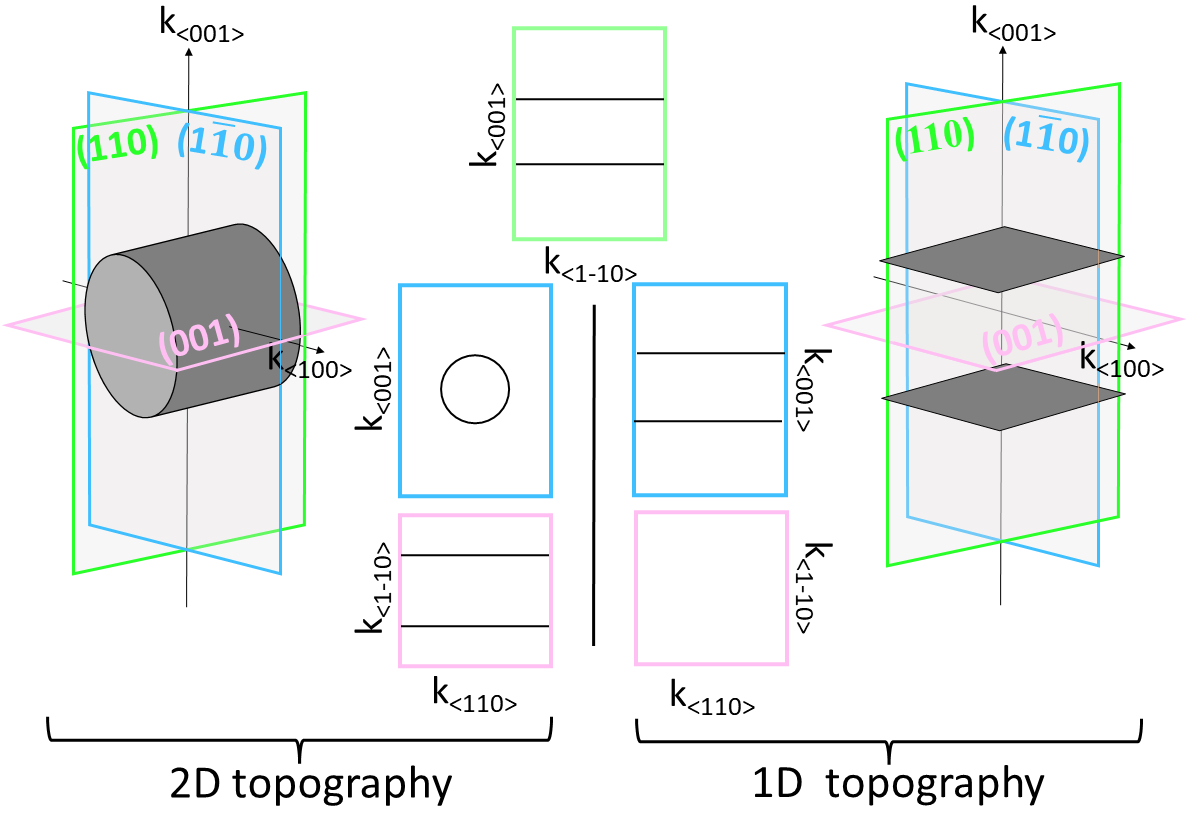}
    \caption{
    		 \footnotesize{
            Schematic drawing of the expected projections of 2D and 1D states on the three planes probed in the ARPES measurements.
    		 }
             %%%
        }
\label{Fig:YNP_FS_topo}

\end{figure}

%%%%%%%%%%%%%%%%%%%%%%%%%%%%%%%%
%%%
%%%%%%%%%%%%%%%%%%%%%%%%%%
%% METHODS SUMMARY
%%%%%%%%%%%%%%%%%%%%%%%%%%
Our ARPES experiments were performed at the CASSIOPEE beamline 
of the Synchrotron SOLEIL (France) and at beamline 2A of the KEK Photon Factory (Japan). Typical energy and angular
resolutions were $15$ meV and $0.25^{\circ}$, respectively. 

In order to generate pristine surfaces for the ARPES experiments, 
high-quality \hkl(110)-oriented YbNi$_4$P$_2$ single crystals were annealed \emph{in-situ}, in UHV conditions, to clean the surface from impurities left after polishing.
Subsequently, the surfaces underwent several \emph{in-situ} cycles, 
of mild bombardment with Ar$^{+}$ ions and annealing to promote recrystallization. The long-range order of the surfaces was verified by means of low-energy electron diffraction (LEED) --see Fig. S1 in [SI].
%% Further details on the sample preparation protocol and on the experimental conditions 
%% of the ARPES measurements can be found in [SI].\\

We compare our ARPES results to DFT calculations where we confine the $f$ electrons of Yb to the core, i.e. we do not consider hybridisation of the $f$ states. We implement this $f$-core configuration in Wien2k \cite{Blaha2019} by calculating the band structure for LuNi$_4$P$_2$ using the lattice parameters of YbNi$_4$P$_2$, with a full $4f$ shell well below the Fermi level $E_F$. LuNi$_4$P$_2$ and YbNi$_4$P$_2$ differ only by one $4f$ hole and thus the band structure calculations of LuNi$_4$P$_2$ result in a band structure where most bands are merely shifted in energy relative to YbNi$_4$P$_2$. The band structure calculations thus reveal the Fermi surfaces and dimensionality of the conduction electrons \cite{Friedemann2008,Friedemann2013b}.
%%%

See the Supplemental Material
%%
% which includes Refs.~\cite{kliemt_crystal_2016,Blaha2019,Krellner2011,Perdew1996,denlinger_comparative_2001,frantzeskakis_hidden_2021,lashell_spin_1996,kirkpatrick_ferromagnetic_2020,manchon_new_2015,ishizaka_giant_2011,di_sante_electric_2013}, 
%%
for additional information about the sample preparation, the experimental methods, 
the reduction and analysis of the ARPES data, and the DFT calculations.

%%%%%%%%%%%%%%%%%%%%%%%%%%
%% RESULTS
%%%%%%%%%%%%%%%%%%%%%%%%%%

In YbNi$_4$P$_2$, the near-$E_F$ $f$-core band structure calculated by density functional theory (DFT) (Fig. S2 in [SI]) shows five different bands dominated by Ni 3$d$ character (Fig. S3 in [SI]), each represented with a different color, straddling the Fermi level. At a binding energy of 80 meV, the energy where we compare the DFT calculations with the experimental results, just three out of the five bands are present and give rise to the three constant energy surfaces represented in Figs. \ref{Fig:YNP_FS}c to \ref{Fig:YNP_FS}e. One can readily observe contours of different dimensionality ranging from quasi-1D (flat sheets) to textbook 3D (doughnut) and of mixed dimensionality (jungle-gym).

In order to experimentally probe the dimensionality of the electronic structure, % predicted by DFT calculations, 
we performed ARPES experiments on YbNi$_4$P$_2$ single crystals in the paramagnetic phase ($T = 16$ K). 
We chose (110)-oriented surfaces  (i.e. the highlighted plane in Fig. \ref{Fig:YNP_FS}b) because this orientation permits us to have spectroscopic fingerprints of all different expected contours. 
%shown in Figs.~\ref{Fig:YNP_DFT}~(c-g). 
%%
Fig. \ref{Fig:YNP_FS}f corresponds to in-plane 
constant energy ($E - E_F = - 80$ meV) cut. The ARPES signal being proportional to the so-called 
photoemission matrix elements, dependent of geometrical considerations 
(orbitals character, sample orientations and light polarization), 
the raw constant energy surface (Fig. S4 in [SI]) shows a strong asymmetry between positive and negative 
wave-vector along $k_{\braket{001}}$~\cite{moser_experimentalists_2017}. 
Using a mirror operation we symmetrized the positive region 
of the raw constant energy cut, corresponding to the resulting surface shown in Fig. \ref{Fig:YNP_FS}f.

\begin{figure}
    \centering
    \includegraphics[width=0.9\linewidth]{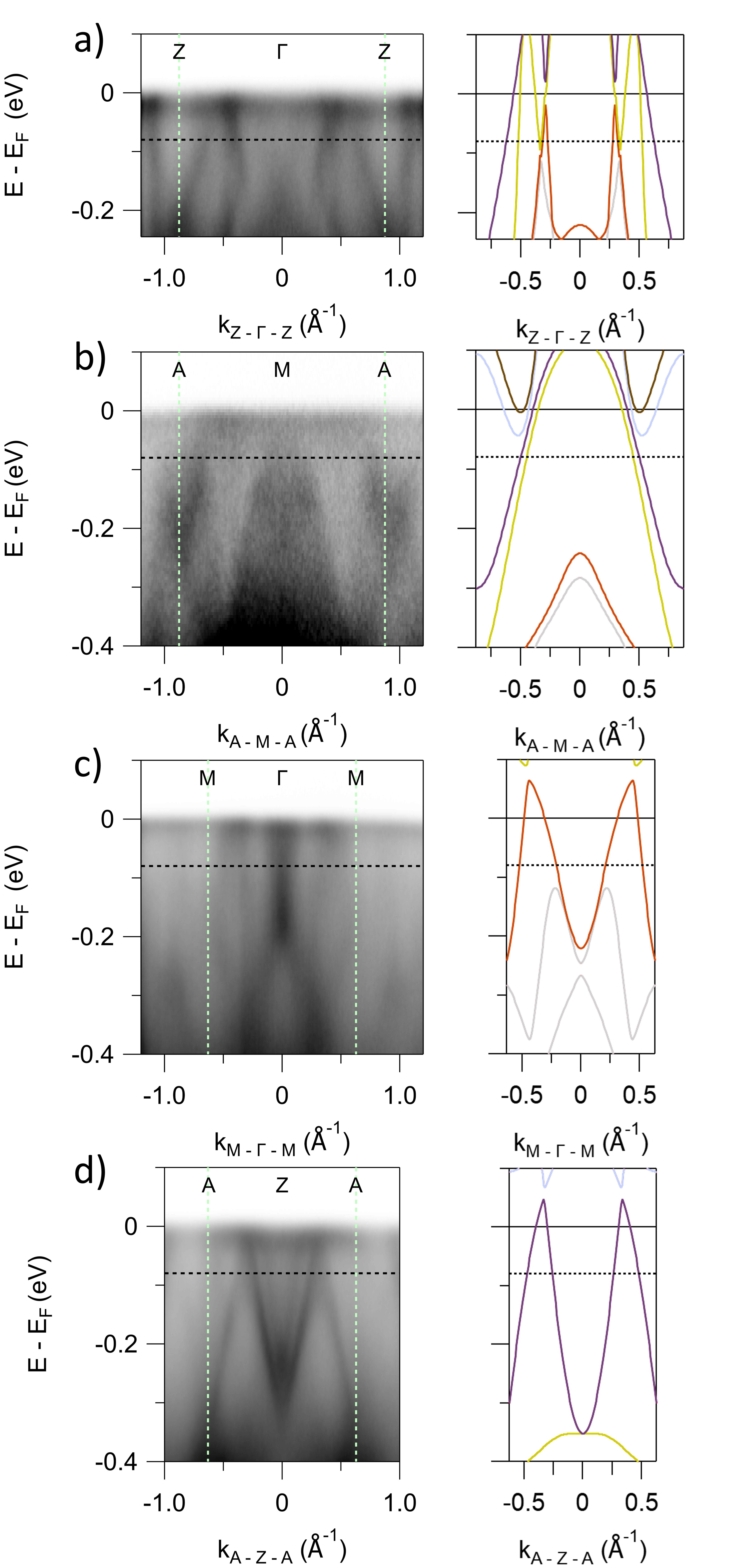}
    \caption{Energy-momentum maps measured along high symmetry lines at $h\nu = 90$ eV 
    		together with their corresponding calculated band structure. 
    		The borders of the DFT plots are the borders of the Brillouin zone and for each ARPES map, the borders of the Brillouin zone are indicated by the dashed light green lines. The black dashed lines indicate the iso-energy line of - 80 meV.
    		(a) measured along Z - $\Gamma$ - Z. The map has been obtained by summing pixel by pixel 
    		the maps measured with horizontally and vertically polarized light, 
    		where the pixel intensity of the maps was divided by the maximal pixel intensity of each map respectively. 
    		(b) measured along A - M - A using horizontally polarized light. 
    		(c) measured along  M - $\Gamma$ - M using vertically polarized light. 
    		(d) measured along A - Z - A using horizontally polarized light.
    		}
    \label{fig:YNP_Ekmaps}
\end{figure}

In the following, we will compare the experimental ARPES results 
shown in Fig. \ref{Fig:YNP_FS}f
with  the theoretical constant energy cut (CEC) (Figs.~\ref{Fig:YNP_FS}c to \ref{Fig:YNP_FS}e) in the measurement plane highlighted 
in Fig.~\ref{Fig:YNP_FS}b, at a constant energy $E - E_F = - 80$ meV. We chose this energy because the experimental contours are best resolved, but later on we will compare the complete energy-momentum dispersion of the states (Fig. \ref{fig:YNP_Ekmaps}). The doughnut is observed in Fig. \ref{Fig:YNP_FS}f as an ellipse centered
around $\Gamma$ with semi-axes along $k_{\braket{1\overline{1}0}}$ of $ 0.48 \pm 0.01 $ \AA$^{-1}$ and $ 0.35 \pm 0.01 $ \AA$^{-1}$ along $k_{\braket{001}}$, which compares well with the theoretical prediction of 0.47 \AA$^{-1}$ and 0.32 \AA$^{-1}$, respectively. Located inside the doughnut and enclosing $\Gamma$, the observed elliptic state is not captured by the calculations and is discussed in Fig. \ref{fig:YNP_Ekmaps}. Flat sheets are observed as a pair of flat stripes parallel to $k_{\braket{1\overline{1}0}}$ at momenta 
$k_{\braket{001}} \approx \pm 0.45$ \AA$^{-1}$ together with small circular pockets at their center ($k_{\braket{1\overline{1}0}} = 0$). 
Lastly, the experimental observation of the jungle-gym is not as straightforward as for the two first CECs. Due to its vicinity to the quasi-1D sheets, it is difficult to disentangle its contribution to the experimental electronic structure. However, the small pockets observed at the A - Z - A edge of the Brillouin zone at a momentum value of $k_{\braket{1\overline{1}0}} \approx 0.2$ \AA$^{-1}$ (green arrow in Fig. \ref{Fig:YNP_FS}f) might be attributed to the jungle-gym despite the relatively poor quantitative agreement with the theoretical contours.
The pockets lying at the A - M - A edge of the Brillouin zone at a momentum value of $k_{\braket{001}} \approx 0.6$ \AA$^{-1}$ do not match with the theoretical computed contours at $E - E_F = - 80$ meV. Nevertheless, they might originate from band 4 or 5 of Fig. S2 in [SI] as their theoretical contours at the Fermi level match satisfactorily the experimental ones at $E - E_F = - 80$ meV. We speculate that the observed difference might be explained by energy shifts not taken into account by standard DFT calculations, for example, energy shifts due to the expected $d - f$ hybridization.
Fig. \ref{Fig:YNP_FS}g shows the calculated contours of the constant energy cut observed in Figs. \ref{Fig:YNP_FS}c to \ref{Fig:YNP_FS}e in the $\braket{001} \times \braket{1\overline{1}0}$ plane. As discussed, the doughnut outer contours compare satisfactorily to the experimental data, while the inner contours are broader than what is experimentally observed. Moreover, the quasi-1D sheets contours are also well captured by the calculations, whereas the jungle-gym contours present some qualitative discrepancies from the measured constant energy cut. Specifically, they fail to describe the exact shape of the pockets along the A - Z - A high-symmetry line, as well as the pockets themselves along the A - M - A high-symmetry line.

Since both $\braket{110}$ and $\braket{1\overline{1}0}$ 
directions are equivalent in tetragonal symmetry, one should expect to observe in the former the same 1D contours as in Fig. \ref{Fig:YNP_FS}f. In order to verify the dimensionality of the experimentally observed quasi-1D sheets, we will now make the distinction of the experimental fingerprints expected of a truly 1D state and of an hypothetical 2D cylindrical-like state that could show a 1D fingerprint in certain planes. Fig. \ref{Fig:YNP_FS_topo} compares the expected experimental fingerprint of 1D and hypothetical 2D cylindrical states in the $\braket{1\overline{1}0}$ x $\braket{001}$ plane, both similar to what is experimentally observed in Fig. \ref{Fig:YNP_FS}f. Therefore, in order to unequivocally determine the dimensionality of the states observed in this plane and confirm their 1D character, it is necessary to probe the $\braket{110}$ x $\braket{001}$ and $\braket{1\overline{1}0}$ x $\braket{110}$ planes, as depicted in Fig. \ref{Fig:YNP_FS_topo}. 

Figure~\ref{Fig:YNP_FS}h shows the out-of-plane constant energy ($E - E_F = - 80$ meV) cut measured by systematically varying the photon energy between 30 and 100 eV (see 3D k-space mapping in [SI] for more details). 
This figure shows, as Fig. \ref{Fig:YNP_FS}f, a pair of identical 1D contours parallel to the $k_{\braket{110}}$ direction  at almost constant momenta $k_{\braket{001}} \approx 0.35$ \AA$^{-1}$. This experimental topography is in full agreement with the scenario of 1D states shown in Fig. \ref{Fig:YNP_FS_topo} (plane with blue contour).

In order to verify that the non-dispersive features observed in Figs.~\ref{Fig:YNP_FS}f and \ref{Fig:YNP_FS}h  are indeed 1D and not artifacts of the ARPES 3D k-space mapping arising from the intrinsic limitations of the out-of-plane mapping in ARPES experiments, we performed a similar photon energy-dependent measurement in a plane where no 1D features are expected (i.e. the $\braket{110}\times\braket{1\overline{1}0}$ plane at the $\Gamma$ point). The measurement is shown in Fig. \ref{Fig:YNP_FS}i and, in line with our reasoning, no sign of 1D contours has been observed at any binding energy. As previously, this experimental topography is in full agreement with the scenario of 1D states shown in Fig. \ref{Fig:YNP_FS_topo} (plane with pink contour).

After comparing the constant energy cuts to the DFT calculations, we now focus on the energy-momentum maps along high symmetry lines. 
Fig. \ref{fig:YNP_Ekmaps} shows a comparison between experimental data and calculated band structure in the $\braket{001}\times\braket{1\overline{1}0}$ plane. 
The band dispersion along the four high-symmetry directions is well captured by the calculations with minor variations with the experimental data.
In Fig. \ref{fig:YNP_Ekmaps}a, along Z - $\Gamma$ - Z, an unexpected electron-like pocket, originating form the most inner ellipse, is observed around $\Gamma$. Moreover, a weakly dispersive band at $E_F$ attributed to the Yb 4$f$ orbitals is also detected. Along the  M - $\Gamma$ - M direction, Fig. \ref{fig:YNP_Ekmaps}c, one can observe spectral weight at the zone center where no state is theoretically predicted, and might be of the same origin as the one observed along Z - $\Gamma$ - Z.
The electron-like pocket in Fig. \ref{fig:YNP_Ekmaps}b, with a band bottom of $\approx$ 175 meV at the A high-symmetry point, is shifted to smaller binding energy compared to the theoretical calculations. 
Finally, in the map measured along the A - Z - A direction, the band originating from the flat sheets Fermi surface at around $-350$ meV is not observed, while band 3 (jungle-gym), crossing the Fermi level, is rigidly shifted to a slightly lower energy (by $\approx$ 100 meV) and may be split due to Rashba-like mechanism. 
A detailed discussion of Fig. \ref{fig:YNP_Ekmaps}d can be found in Fig. S6 in [SI]. We stress that despite these quantitative discrepancies, the overall shape and qualitative dispersion of the theoretical bands stay in good agreement with the experimental features as already demonstrated during the discussion of Fig. \ref{Fig:YNP_FS}f. Most importantly, our data unequivocally prove the existence of quasi-1D states and states of higher dimensionality in the electronic structure of YbNi$_4$P$_2$.

%%%%%%%%%%%%%%%%%%%%%%%%
% DISCUSSION / CONCLUSION
%%%%%%%%%%%%%%%%%%%%%%%%
In conclusion, our results provide new insights on the electronic structure of YbNi$_4$P$_2$ 
and on the link between dimensionality and ferromagnetic quantum-criticality. We have experimentally shown that, while YbNi$_4$P$_2$ hosts quasi-1D states, 
part of its electronic structure is also of higher dimensionality. 
Therefore, our results suggest that a purely 1D electronic structure is not a necessary condition 
for FM QC. This conclusion is in line with recent ARPES results on CeRh$_6$Ge$_4$, a metal hosting a pressure-induced QCP \cite{Kotegawa2019, shen_strange-metal_2020}. In this pure ferromagnetic Kondo lattice system, no 1D features are observed
in its electronic structure~\cite{wu_anisotropic_2021}, while transport studies observed hallmarks of a FM quantum critical point~\cite{wang_localized_2021}.
These studies further suggested that the occurrence of FM QC
was ascribed to the anisotropic hybridization interaction between the conduction
and Ce 4$f$ bands~\cite{wu_anisotropic_2021}. Moreover, in the case of YbNi$_4$P$_2$, 
our data reveal that electronic correlations, $d - f$ hybridization and spin-orbit coupling 
are all essential ingredients to describe its electronic structure 
and need to be worked out more conclusively, using Dynamical Mean Field Theory (DMFT) or extended DMFT to incorporate Kondo physics \cite{peters_spin-selective_2012,grempel_locally_2003,jang_one-dimensionality_2023}.

%%%%%%%%%%%%%%%%%%%%%%%%%%%%%%%%%%%%%%%%%%%%%%%%%%%%%%%%%%%%%%%%%%%%%%%%%%%%%%%%%%%%%%%
%%%%%%%%%%%%%%%%%%%%%%%%%%%%%%%%%%%%%%%%%%%%%%%%%%%%%%%%%%%%%%%%%%%%%%%%%%%%%%%%%%%%%%%
%%%%%%%%%%%%%%%%%%%%%%%%%%%%%%%%%%%%%%%%%%%%%%%%%%%%%%%%%%%%%%%%%%%%%%%%%%%%%%%%%%%%%%%
\acknowledgments 
%% We thank XYZ for discussions.
We thank F.~Bertran for technical support during the experiments at SOLEIL. 
Work at ISMO was supported by public grants from the French National Research Agency (ANR), 
projects Fermi-NESt No. ANR-16-CE92-0018 and SUPERNICKEL No. ANR-21-CE30-0041-05, 
and by the CNRS International Research Project EXCELSIOR. 
Work at Tohoku University was supported by Grants-in-Aid for Scientific Research (Nos. 22H01948 and 20KK0117) 
from the Japan Society for the Promotion of Science. 
Experiments at KEK-PF were performed under the approval of the Program Advisory Committee 
(Proposals 2022G515 and 2021S2-002) at the Institute of Materials Structure Science at KEK. 
Work at the University of Bristol was supported by the EPSRC through grant EP/L015544/1. 
The authors gratefully acknowledge support by the Deutsche Forschungsgemeinschaft (DFG) 
through Grants No. KR 3831/4-1 and TRR288 (No. 422213477, Project No. A03).
%%%%%%%%%%%%%%%%%%%%%%%%%%%%%%%%%%%%%%%%%%%%%%%%%%%%%%%%%%%%%%%%%%

%%%%%%%%%%%%%%%%%%%%%%%%%%%%%%%%
\section{SUPPLEMENTAL MATERIAL}
%%%%%%%%%%%%%%%%%%%%%%%%%%%%%%%%

%%%%%%%%%%%%%%%%%%%%%%%%%%%%%%%
\subsection{Crystal growth}
%%%%%%%%%%%%%%%%%%%%%%%%%%%%%%%
Single crystals of YbNi$_4$P$_2$ were grown from a levitating melt using the Czochralski method under an Ar overpressure of 20 bar according to the procedure described in \cite{kliemt_crystal_2016}.
The crystal structure was characterized by powder x-ray diffraction on crushed single crystals, using Cu-K$_{\alpha}$ radiation on a Bruker D8 diffractometer and confirmed the tetragonal ZrFe$_4$Si$_2$ structure type ($P4_2/mnm$) with lattice parameters $a=7.0560(3)$\,\AA, and $c=3.5876(5)$\,\AA. Energy-dispersive x-ray spectroscopy revealed the stoichiometry of the 142 compound within an error of 2\,at\%. The single crystals were oriented by Laue method using the white radiation of a tungsten anode (M\"uller Mikro 91) and afterwards cut using a spark erosion device. In preparation of the ARPES measurements the (110) surface of the crystals were polished down to a surface roughness of $\approx 0.01\,\mu m$ with a precision of orientation $<1^{\circ}$ (MaTeck).

%%%%%%%%%%%%%%%%%%%%%%%%%%%%%%%
\subsection{ARPES measurements}
%%%%%%%%%%%%%%%%%%%%%%%%%%%%%%%
%%%%%%%%%%%%%%%%%%%%%%%%%%
%FIGURE LEED
%%%%%%%%%%%%%%%%%%%%%%%%%
%%%%%%%%%%%%%%%%%%%%%%%%%%%%%%%
\begin{figure}[t!h]
	\centering
       \includegraphics[clip, width=0.4\linewidth]{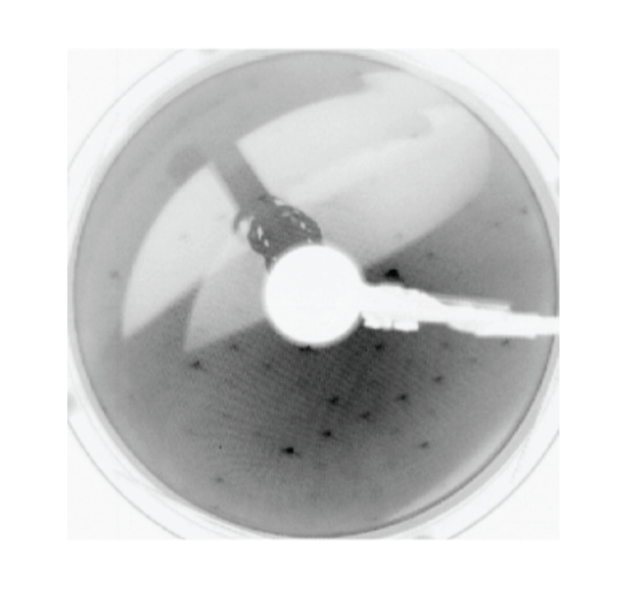}
   \caption{\footnotesize{
       	 LEED image of YbNi$_4$P$_2$ 
       	 obtained after in-situ annealing, right before ARPES measurements.
       	 }
            %%%
       }
\label{Fig:LEED}
\end{figure}
%%%%%%%%%%%%%%%%%%%%%%%%%%%%%%%
%%

%%
Typical electron energy and angular resolutions were 15~meV and 0.25$^{\circ}$. 
In order to generate pristine surfaces for the ARPES experiments, 
the single crystals of YbNi$_4$P$_2$ were subject to various cycles
of soft Ar bombardment (800 eV at a grazing angle of $30^{\circ}$ for about 20 minutes)
followed by annealing for 5-10 minutes 
at approximately 550-600$^{\circ}$C in UHV conditions. 
Low-energy electron diffraction (LEED) 
was employed to verify the long-range crystallinity and cleanliness of our surfaces 
after preparation --see Fig. \ref{Fig:LEED}.
Clean surface preparation and ARPES experiments were performed at the CASSIOPEE beamline of Synchrotron SOLEIL (France) 
and at beamline 2A of KEK-Photon Factory (KEK-PF, Japan) 
using hemispherical electron analyzers with vertical and horizontal slits, respectively. 
%%

%%%%%%%%%%%%%%%%%%%%%%%%%%%%%%%%%%
\subsection{3D $k$-space mapping}
%%%%%%%%%%%%%%%%%%%%%%%%%%%%%%%%%%
Within the free-electron final state model, ARPES
measurements at constant photon energy $h\nu$ give the electronic
structure at the surface of a spherical cap of radius
$k = \sqrt{\frac{2m_{e}}{\hbar^{2} } }(h\nu - \Phi + V_{0})^{1/2}$
. Here, $m_{e}$ is the free electron
mass, $\Phi$ is the work function, and $V_{0}$ = 11 eV is the
inner potential of YbNi$_{4}$P$_{2}$. Measurements around normal
emission provide the electronic structure in a plane
nearly parallel to the surface plane. Likewise, measurements
as a function of photon energy provide the electronic
structure in a plane perpendicular to the surface.

%%%%%%%%%%%%%%%%%%%%%%%%%%%%%%%%%%%%%%%%%%%%%%%%%
\subsection{Density functional theory calculations}
\begin{figure*}
	\centering
        \includegraphics[clip, width=\linewidth]{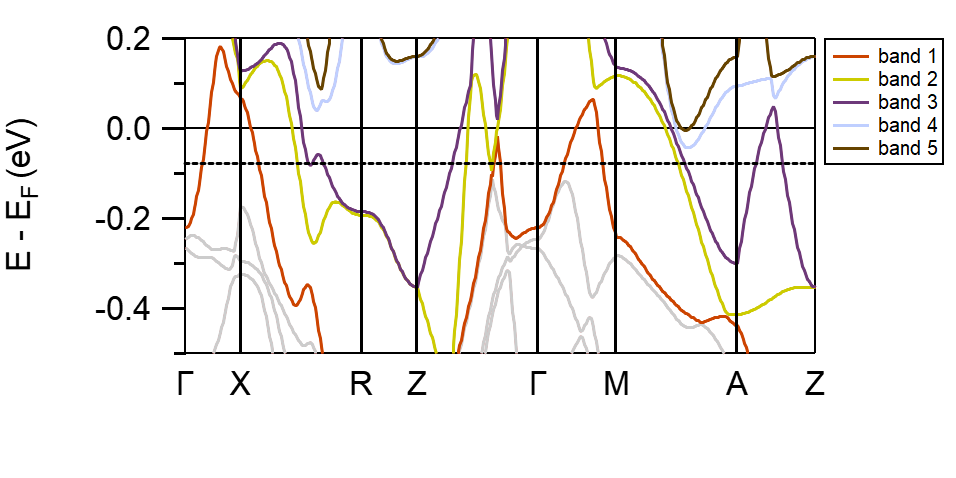}
    \caption{
    		 \footnotesize{
    		 f-core YbNi$_4$P$_2$ band structure obtained by DFT calculations. 
    		 Each band crossing the Fermi level is represented in a different color. The gray lines are bands not crossing the Fermi level. The black dashed line indicates the iso-energy line of - 80 meV.
    		 }
             %%%
        }
\label{Fig:YNP_spaghetti}

\end{figure*}

%%%%%%%%%%%%%%%%%%%%%%%%%%%%%%%%%%%%%%%%%%%%%%%%%
Band structure calculations of the $f$-core configuration have been implemented in Wien2k \cite{Blaha2019} by calculating the band structure of LuNi$_4$P$_2$ using the experimentally established crystal structure of YbNi$_4$P$_2$ \cite{Krellner2011}. The exchange correlation potential is approximated using the Perdew-Burke-Ernzerhof generalized gradient \cite{Perdew1996}. We use RKmax=9 resulting in a total-energy convergence better than 10 meV and an energy range -5 to 5 Ry thus treating the 4f, 5s, 5p, 5d, and 6s electrons of Lu and the 3d and 4s as well as the 3s and 3p states of Ni and P, respectively as valence states. Spin-orbit interactions are approximated using relativistic local orbits with a p$_{1/2}$ radial basis for Lu and scalar-relativistic orbits for Ni. The total energy converges self-consistently better than 1 meV using 3480 points in the Brillouin zone. The band structure is then evaluated for selected paths and planes in reciprocal space on a fine mesh for comparison with ARPES results.
Fig.~\ref{Fig:YNP_spaghetti} shows the near-$E_F$ f-core band structure of YbNi$_4$P$_2$ calculated by density-functional-theory (DFT). Five different bands dominated by Ni 3$d$ character (Fig. \ref{Fig:orbital_character}), each represented with a different color, straddle the Fermi level. At a binding energy of 80 meV, the energy where we compare the DFT calculations with the experimental results, just three out of the five bands are existing and give rise to the three constant energy surfaces represented in Fig. 1c to 1e in the main text.

%%%%%%%%%%%%%%%%%%%%%%%%%%%%%%%%%%%%%%%%%%%%%%%%%%%%%%%%%%

%%%%%%%%%%%%%%%%%%%%%%%%%%%%%%%%%%%%%
\subsection{Orbital character at $E_F$}
\begin{figure*}[h]
	\centering
        \includegraphics[clip, width=0.65\linewidth]{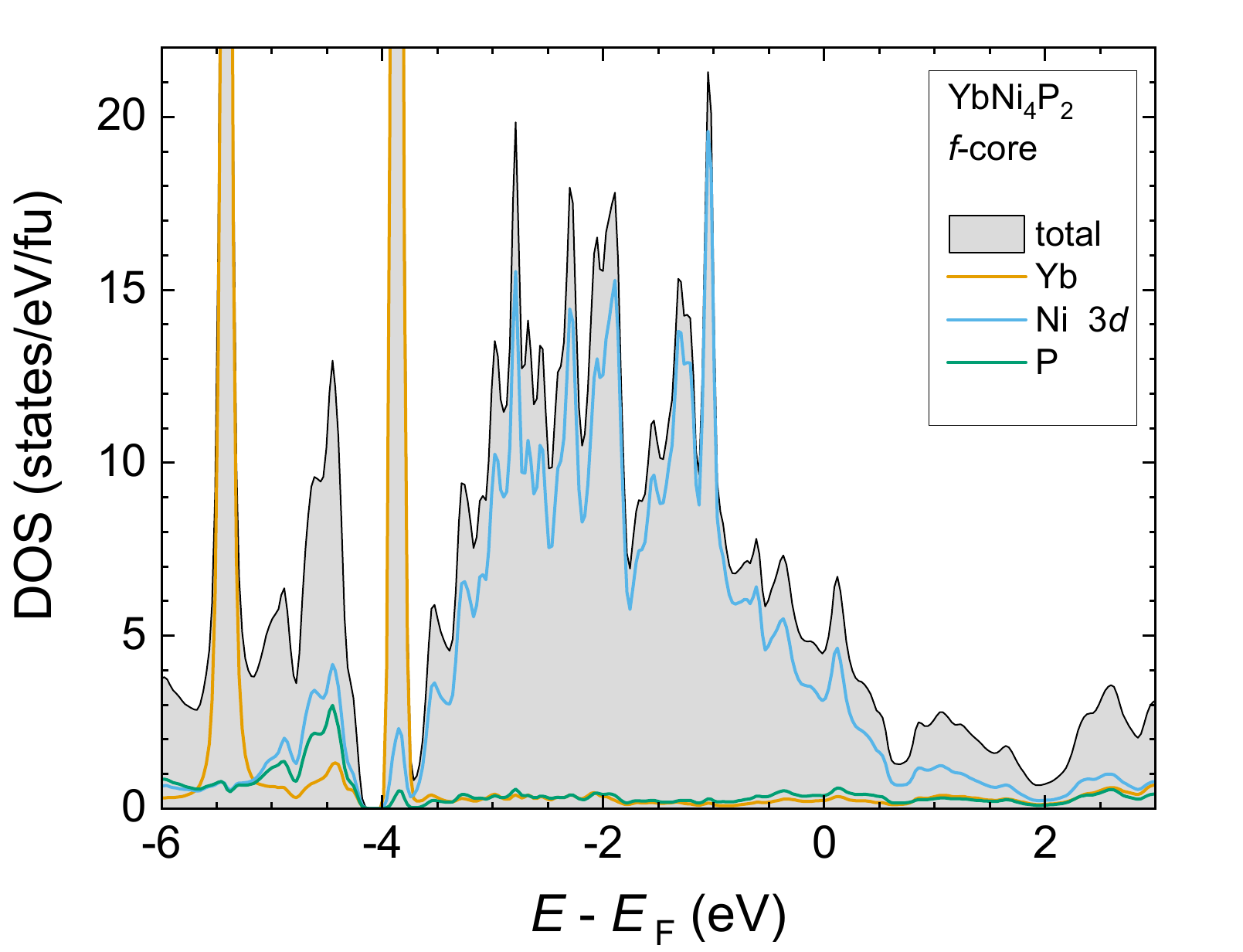}
    \caption{
    		 \footnotesize{
                Density of States (DoS) of $f$-core YbNi$_4$P$_2$. The total DoS is represented by the gray-shaded area under the black curve, while the contributions of Yb, Ni and P to the DoS are represented by the orange, blue and green curve, respectively. 
        	 }
             %%%
        }
    \label{Fig:orbital_character}
\end{figure*}

%%%%%%%%%%%%%%%%%%%%%%%%%%%%%%%%%%%%%
The states near the Fermi energy are dominated by Ni-$3d$ character as can be seen from Fig. \ref{Fig:orbital_character}. Only small contributions from Yb and P are present in the $f$-core calculation at the Fermi energy. Yet, the magnetism of YbNi$_4$P$_2$ is not of nickel character \cite{Krellner2011}. Instead, the magnetism stems from the Yb $f$ moments which also contribute to the Density of States (DoS) at low temperatures via the Kondo effect (not captured by the $f$-core DFT calculation). 
%%%%%%%%%%%%%%%%%

%%%%%%%%%%%%%%%%%%%%%%%%%%
\subsection{Raw ARPES data}
\begin{figure}
	\centering
        \includegraphics[clip, width=0.6\linewidth]{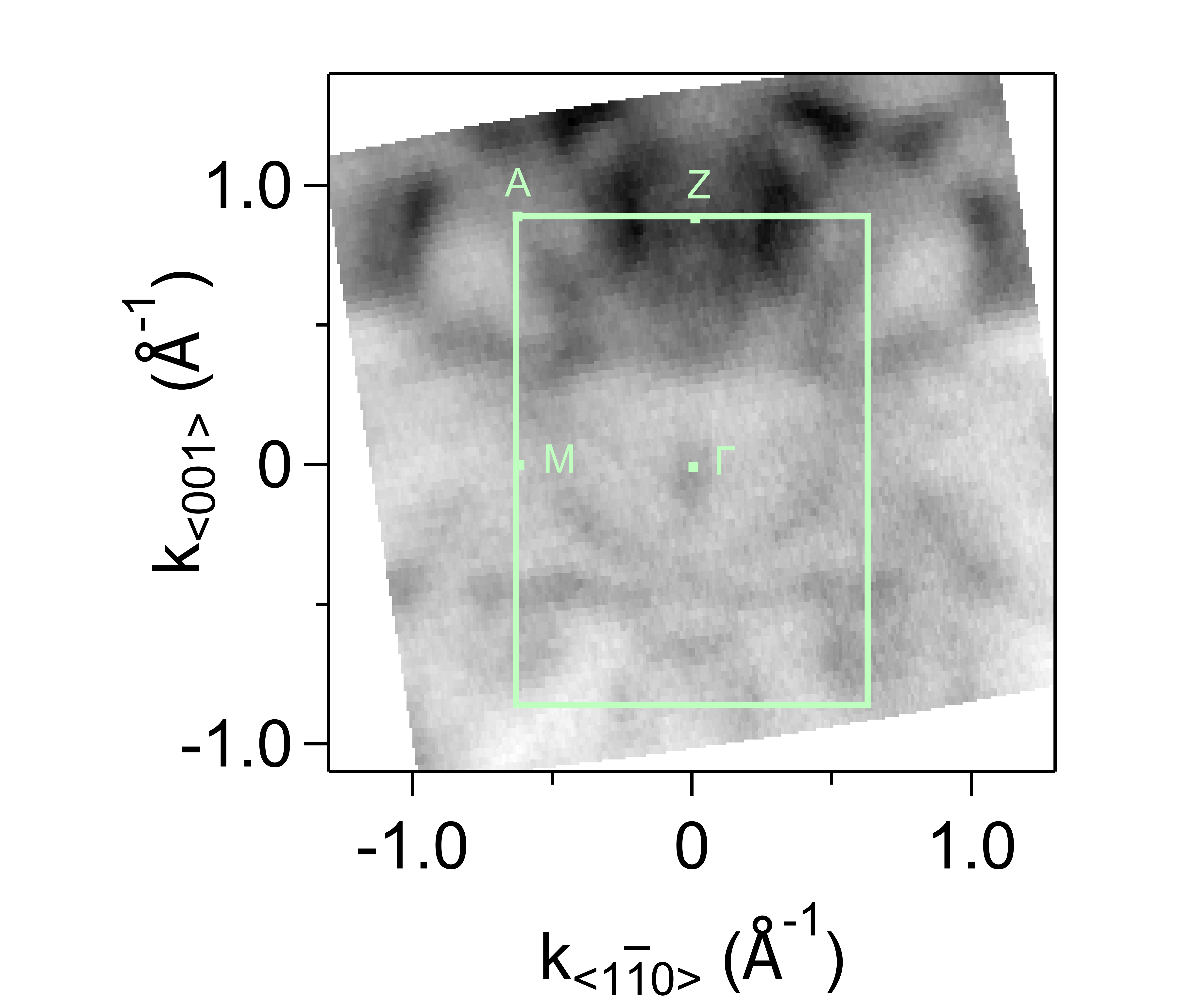}
    \caption{\
    		 \footnotesize{Constant energy cut at $E - E_F$ = - 80 meV in the $\braket{1\overline{1}0}$ x $\braket{001}$ plane, measured with LH polarization and a photon energy of 90 eV.
        	 }
             %%%
        }
\label{Fig:raw_ARPES}
\end{figure}

%%%%%%%%%%%%%%%%%%%%%%%%%%
The photoemission matrix elements are responsible of an asymmetry between the positive and negative parts along $\braket{001}$ for the in-plane constant energy surface (CES) in the $\braket{1\overline{1}0}$ x $\braket{001}$ plane (Fig. 1f of the main text). Hence, as explained in the main paper, we symmetrized the data by a mirror operation and showed the negative part k$_{<001>} < 0$. Fig. \ref{Fig:raw_ARPES} shows the raw in-plane CES in the $\braket{1\overline{1}0}$ x $\braket{001}$ plane at $E - E_F$ = - 80 meV.

%%%%%%%%%%%%%%%%%%%%%%%%%%%%%%%%%%%%
\subsection{In-plane Fermi surface}
\begin{figure}
	\centering
        \includegraphics[clip, width=\linewidth]{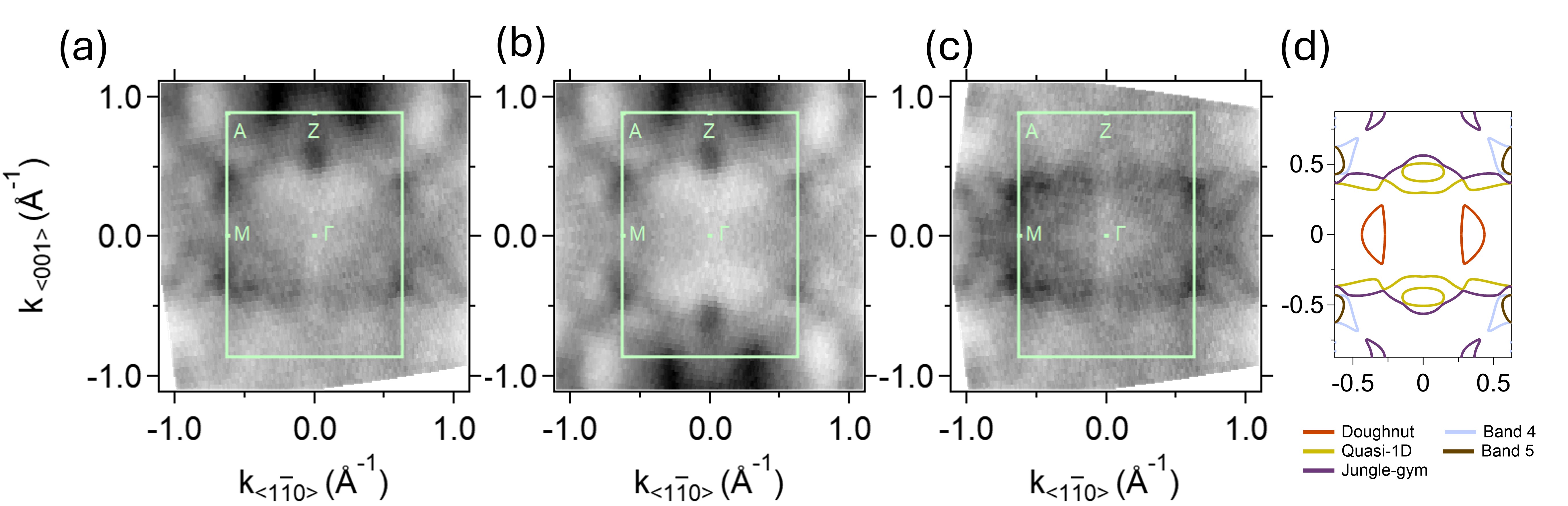}
    \caption{\
    		 \footnotesize{(a) Constant energy cut at $E - E_F$ = 0 eV in the $\braket{1\overline{1}0}$ x $\braket{001}$ plane, measured with LH polarization and a photon energy of 90 eV. (b) Same data but symmetrized along the $k_{\braket{001}} = 0$ line for  $k_{\braket{001}}$ negative. (c) Same as (b) but for $k_{\braket{001}}$ positive. For (a), (b) and (c) the Brillouin zone and high symmetry points are superimposed on the data. (d) Calculated constant energy cut at $E - E_F$ = 0 eV in the first Brillouin zone. 
        	 }
             %%%
        }
\label{Fig:FS_at_EF}
\end{figure}

%%%%%%%%%%%%%%%%%%%%%%%%%%%%%%%%%%%%
The constant energy cut at $E - E_F$ = 0 meV (integrated 40 meV bellow $E_F$) is shown in Fig. \ref{Fig:FS_at_EF}a. Due to strong matrix elements effects, the photoemission signal is not symmetric in the $k_{\braket{001}}$ negative and $k_{\braket{001}}$ positive part of the image. Fig. \ref{Fig:FS_at_EF}b and \ref{Fig:FS_at_EF}c are, respectively, the symmetrized constant energy cuts of the positive and negative part of Fig. \ref{Fig:FS_at_EF}a. We will focus on these figures to discuss the observed features and compare them to the theoretical Fermi surface of Fig. \ref{Fig:raw_ARPES}d. First of all, one can readily observe in paenl (a) an ensemble of quasi-flat states straddling the Brillouin zone parallel to the $k_{\braket{1\overline{1}0}}$ direction at an approximate $k_{\braket{001}}$ value ranging from 0.4 to 0.5 \AA$^{-1}$. 

Through comparison to DFT calculations at $E_F$ shown in panel (d), we can assign the quasi-flat states located in the $k_{\braket{001}}$ positive region, best seen in panel (c) at momentum value $k_{\braket{001}} \approx 0.5$ \AA$^{-1}$, to the jungle-gym FS. The quasi-1D states observed in the $k_{\braket{001}}$ negative region are fingerprints of the flat sheets FS, best seen in panel (c) at momentum value $k_{\braket{001}} \approx 0.4$ \AA$^{-1}$. 
In panel (c), between the two pairs of the aforementioned bands, we can see intensity features shaped as half circles that we assign to the doughnut contour. Finally in panel (b), at the edge of the A - Z - A line, one can identify the feature in the middle of A - Z as originating from the jungle-gym. \\
Since DFT is not able to capture the experimental FS at $E_F$ because of the Kondo hybridisation, only qualitative comparison with the experimental data is possible. Therefore, while we are able to track and identify the different features making the complete FS at $E_F$, we are not able to extract the exact positions and Fermi radii of these states. In the main text (Fig. 3), one may notice diffuse non-dispersing intensity near the Fermi energy along the main high symmetry directions. This photoemission intensity of presumably $f$-origin interacts with the rest of the states altering their energy-momentum dispersion. This is clearly shown in the next subsection when we discuss the electronic structure along A - Z - A, and may as well happen for the doughnut Fermi contours. Moreover, the doughnut Fermi contours are in the immediate vicinity of $\Gamma$ where an electron pocket with no theoretical counterpart has been observed in Figs. 3a and 3c. For these reasons, the theoretical Fermi surface cannot capture all experimental details.
%%%%%%%%%%%%%%%%

%%%%%%%%%%%%%%%%%%%%%%%%%%%%%%%%%%%%%%%%%%%%%%%%%%%%%%%
\subsection{Details on the electronic structure along A - Z - A}
%%%%%%%%%%%%%%%%%%%%%%%%%%%%%%%%%%%%%%%%%%%%%%%%%%%%%%%
%%%%%
% FIGURE 3
%%%%%
%%%%%%%%%%%%%%%%%%%%%%%%%%%%%%%%
\begin{figure}
	\centering
        \includegraphics[clip, width=\linewidth]{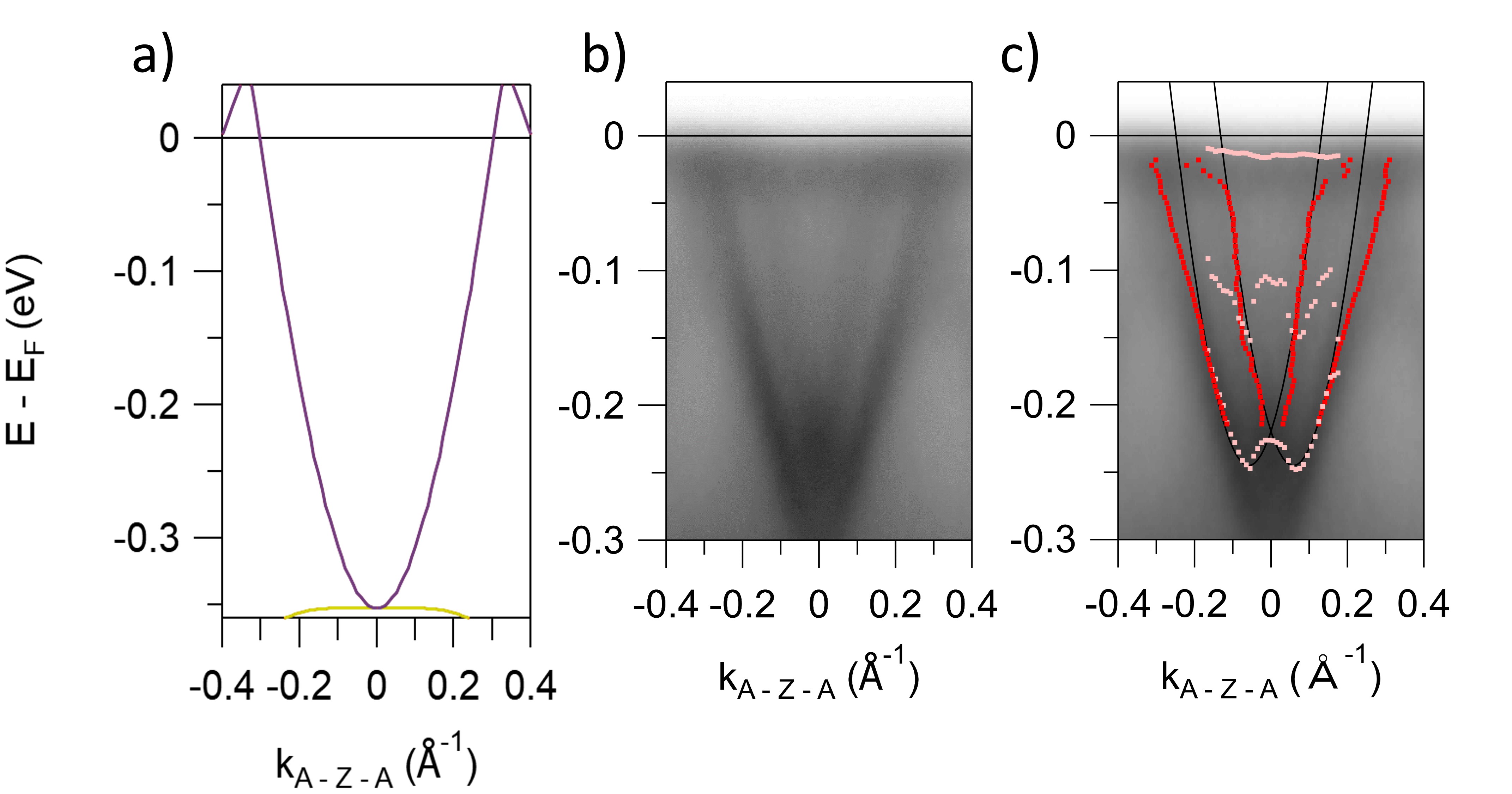}
    \caption{\
    		 \footnotesize{(a) Calculated energy-momentum map along the A - Z - A direction. 
    		 The colored bands correspond to bands of Fig. \ref{Fig:YNP_spaghetti}.
            (b) Energy-momentum map along A - Z - A measured with linear horizontal polarized light.
    		 (c) Fit of the two parabolic bands in panel (b) using the procedure described in the text. Red (pink) markers are extracted from the MDCs (EDCs) fits, respectively.  
        	 }
             %%%
        }
\label{Fig:Rashba}
\end{figure}
%%%%%%%%%%%%%%%%%%%%%%%%%%%%%%%%
%%%
Other interesting features of the experimental band structure can be seen in the energy-momentum dispersion map along A - Z - A, Fig. 3d in the main text. DFT calculations shown in Fig.~\ref{Fig:Rashba}a, predict that the electronic state originating from the jungle-gym contour crosses the Fermi level two times. Our experimental results (Fig.~\ref{Fig:Rashba}b) show that the two inner branches of this state appear to be split resulting in four crossing points at the Fermi level. Moreover, one can observe the presence of spectral weight right below $E_F$ of Yb 4f origin, not captured by DFT. This is reminiscent of other heavy fermion systems where strong interaction between localized $f$ orbitals and other dispersive sates results in the formation of heavy electron bands just below $E_F$ \cite{denlinger_comparative_2001,frantzeskakis_hidden_2021}. In order to obtain more information on the near-$E_F$ electronic structure we have performed further analysis of the experimental data. Figure ~\ref{Fig:Rashba}c presents a more detailed study of the splitting band structure near the Fermi level. Each red marker is obtained by fitting either the energy distribution curves(EDC) or the molecular distribution curves (MDC) and taking the local maxima. This procedure is applied for every EDC between $\pm 0.17$ \AA $^{-1}$ with an integration range of 0.01 \AA$^{-1}$ and MDC between 0.02 eV and 0.22 eV with an integration range of 0.004 eV. In this way, we can capture the central part of the flat band just below $E_F$, but also two parabolic bands that cross each other at a high symmetry point with clear sign of $d-f$ hybridization. We fit each of the two parabolic branches with a cosine function which was also used to fit the calculated parent band with great accuracy. There are clear signs of a strong hybridization between this set of parabolic states and the flat $f$-states. Both inner and outer branches of the conducting state deviate substantially from the predicted dispersion (from the fit) to higher k-values, i.e. undergo an effective mass enhancement, as they approach the energy position of the flat $f$-state

Because of the symmetry breaking along the [110] direction, a potential gradient in the out-of-plane direction confines this parabolic state, and therefore, it may be prone to the Rashba effect when spin-orbit coupling comes into play \cite{lashell_spin_1996}. 
This could be an important observation as the Rashba effect is a direct effect of enhanced spin-orbit coupling and the latter has been related to a possible mechanism for the existence of a ferromagnetic quantum critical point in systems with higher dimensionality \cite{kirkpatrick_ferromagnetic_2020}. It is therefore worthwhile to evaluate the magnitude of a potential Rashba effect in YbNi$_4$P$_2$, although the aforementioned mechanism cannot be directly applied to this case due to the centrosymmetric nature of the crystal. Considering the hypothesis of a Rashba splitting, the magnitude of the Rashba effect can be quantified by the Rashba parameter $\alpha_R$ which is related to the energy difference of the two branches at the band extrema. From our fit, we could estimate a Rashba parameter of $\approx$ 0.9 eV\AA. The value of the $\alpha_R$ Rashba parameter for YbNi$_4$P$_2$ is much larger than the typical values of most semiconductors \cite{manchon_new_2015}, and is around one sixth of the values in giant Rashba effect bulk crystals \cite{ishizaka_giant_2011,di_sante_electric_2013}.\\

The electronic structure of YbNi$_4$P$_2$ includes conduction states of low-dimensionality: there are states subject to in-plane confinement due to the presence of the surface, or even to extra confinement within the surface plane as in the case of the 1D sheets. Moreover, YbNi$_4$P$_2$ contains $f$-electrons with strong magnetic moments which interact with the conduction states. A surface potential gradient, the presence of magnetic moments and an appreciable $d-f$ hybridisation are all ingredients that favour the development of Rashba-type spin-splitting of the conduction states.

%%%%%%%%%%%%%%%%%%%%%%%%
%% \bibliographystyle{unsrtnat}
%% \bibliography{YNP_bib}

%%%%%%%%%%%%%%%%%%%%%%%
\end{document}